\documentclass[aps,prl,groupaddress,showpacs,twocolumn]{revtex4}
\usepackage{amsfonts}
\usepackage{bm}
\usepackage{epsfig,amsmath,graphicx,amssymb,overpic}
\usepackage{color}

\def\be{\begin{equation}}
\def\ee{\end{equation}}
\def\bee{\begin{eqnarray}}
\def\ene{\end{eqnarray}}
\def\bes{\begin{subequations}}
\def\ees{\end{subequations}}

\begin{document}

\title{Spontaneous Parity--Time Symmetry Breaking and Stability of Solitons
in Bose-Einstein Condensates}
\author{Zhenya Yan$^{1}$}
\author{Bo Xiong$^{2}$}
\email{stevenxiongbo@gmail.com}
\author{W. M. Liu$^{2}$}
\affiliation{$^{1}$Key Laboratory of Mathematics Mechanization, Institute of Systems
Science, AMSS, Chinese Academy of Sciences, Beijing 100190, China \\
$^{2}$Beijing National Laboratory for Condensed Matter Physics, Institute of
Physics, Chinese Academy of Sciences, Beijing 100190, China}
\date{\today }

\begin{abstract}
We report explicitly a novel family of exact $\mathcal{PT}$-symmetric
solitons and further study their spontaneous $\mathcal{PT}$ symmetry
breaking, stabilities and collisions in Bose-Einstein condensates trapped in
a $\mathcal{PT}$-symmetric harmonic trap and a Hermite-Gaussian gain/loss
potential. We observe the significant effects of mean-field interaction by
modifying the threshold point of spontaneous \textit{$\mathcal{PT}$}
symmetry breaking in Bose-Einstein condensates. Our scenario provides a
promising approach to study $\mathcal{PT}$-related universal behaviors in
non-Hermitian quantum system based on the manipulation of gain/loss
potential in Bose-Einstein condensates.
\end{abstract}

\maketitle

Parity ($\mathcal{P}$) and time-reversal ($\mathcal{T}$) symmetries are
fundamental notions in physics. There has already been much attention to
these systems which admire the combined $\mathcal{PT}$ symmetry but do not
obey $\mathcal{P}$ and $\mathcal{T}$ symmetries separately. Despite the fact
that $\mathcal{PT}$-Hamiltonian can, in general, be non-Hermitian, the
pioneering work of Bender and Boettcher~\cite{Bender98} showed that a family
of non-Hermitian Hamiltonian with $\mathcal{PT}$ symmetry, $\hat{H}=\hat{p}+%
\hat{x}^{2}(i\hat{x})^{\epsilon }$, can still show entirely real spectra
whereas the generated dynamics is (pseudo) unitary corresponding to the
\textit{unbroken} $\mathcal{PT}$ symmetry. Their results have inspired
considerable scientific attention devoted to study of non-Hermitian $%
\mathcal{PT}$-symmetric Hamiltonian in many aspects, such as $\mathcal{PT}$%
-symmetric classical-mechanics theory~\cite{CMT}, $\mathcal{PT}$-symmetric
quantum mechanics~\cite{CQM}, $\mathcal{PT}$-symmetric quantum field theory~%
\cite{QM}, pseudo-Hermitian quantum mechanics~\cite{PQM}, Lie algebras~\cite%
{LA}, complex crystal~\cite{CC} and $\mathcal{PT}$-symmetric wave chaos~\cite%
{chaos} etc. Some review papers (\cite{Bender07}, for example) present an
overview of the theoretical and experimental progress on the study of
non-Hermitian $\mathcal{PT}$-symmetric Hamiltonian.

For the case considered here, given that the action of the parity $\mathcal{P%
}$ is linear and has the effect $\hat{p}\rightarrow -\hat{p}$ and $\hat{x}%
\rightarrow -\hat{x}$ whereas the time operator $\mathcal{T}$ is antilinear
and has the effect $\hat{p}\rightarrow -\hat{p}$, $\hat{x}\rightarrow \hat{x}
$, and $i\rightarrow -i$, it then follows that a necessary (but not
sufficient) condition for a Hamiltonian to be $\mathcal{PT}$ symmetric is $V(%
\hat{x})=V^{\ast }(-\hat{x})$. In other words, $\mathcal{PT}$ symmetry
implies that the real and imaginary parts of the $\mathcal{PT}$-symmetric
potential $V(\hat{x})\equiv V_{\mathrm{R}}(\hat{x})+iV_{\mathrm{I}}(\hat{x})$
should be spatially symmetric and antisymmetric, respectively, i.e., $V_{%
\mathrm{R}}(\hat{x})=V_{\mathrm{R}}(-\hat{x})$ and $V_{\mathrm{I}}(\hat{x}%
)=-V_{\mathrm{I}}(-\hat{x})$. It turns out that the $\mathcal{PT}$-symmetric
condition is sufficient to guarantee that the energy spectrum is real and
time evolution is unitary, the condition of Dirac Hermiticity is not
necessary~\cite{Bender07}.

There have been several experimental and theoretical studies on
one-dimensional Schr\"{o}dinger equation with different $\mathcal{PT}$%
-symmetric potentials such as the periodic potentials~\cite{Period},
non-periodic potentials~\cite{Nonp} and $\mathcal{PT}$-symmetric lattices
\cite{PTL}. Some experiments in the framework of wave optics based on $%
\mathrm{Al}_{x}\mathrm{Ga}_{1-x}$~\cite{Exp1} and Fe-doped $\mathrm{LiNbO}%
_{3}$~\cite{Exp2} provide experimental observation of the behaviors in the
both passive and active $\mathcal{PT}$-symmetric optical coupled two-channel
systems. More recently, Musslimani \textit{et al.}~\cite{Muss} have shown
that the optical nonlinearity can shift the $\mathcal{PT}$-symmetric
threshold and in turn allow nonlinear eigenmodes with real eigenvalues to
exist in the Scarff II potential~\cite{ZA} and periodic potentials~\cite{PP}.

Very interestingly, by using the recently realized experimental techniques
\cite{Dis}, $\mathcal{PT}$-symmetric gain/loss potential is very promising
to be experimentally realizable with BECs in the near future. Moreover, a
dissipative source (non-$\mathcal{PT}$-symmetric potentials, i.e., the
gain/loss distribution chosen as a Gaussian function) have been used to
generate nonlinear coherent excitations in Bose-Einstein condensates (BECs)
\cite{Dis}. As is well known that, for ultra-cold alkali atoms, the
confining potential is usually well approximated with the quadratic form~%
\cite{BEC}. Thus, the natural question appeared is whether the spontaneous $%
\mathcal{PT}$ symmetry breaking and stability of solitons are universal. The
goal of this letter is to give the positive answer to this question by
showing that $\mathcal{PT}$-symmetric soliton-type solutions are pretty
natural in BECs trapped by a harmonic potential and synthetic $\mathcal{PT}$%
-symmetric gain/loss potential.

In this Letter, we find a new class of exact $\mathcal{PT}$-symmetric
solitons and further study their spontaneous $\mathcal{PT}$ symmetry
breaking, stabilities and collisions in BECs with a $\mathcal{PT}$-symmetric
harmonic trap and Hermite-Gaussian gain/loss potential. In particular, the
mean-field interaction drastically modifies the threshold point of
spontaneous \textit{$\mathcal{PT}$} symmetry breaking of solitons in BECs.
These results provide a promising approach towards observation of other $%
\mathcal{PT}$-related universal features in new classes of $\mathcal{PT}$%
-synthetic system.

We concentrate on a quasi-one-dimensional (1D) BECs in the mean-field regime
described by the 1D Gross-Pitaevskii (GP) equation with complex $\mathcal{PT}
$-symmetric potential
\begin{equation}
i\hbar \frac{\partial \Psi }{\partial t}\!=\!\left[ -\frac{\hbar ^{2}}{2m}%
\frac{\partial ^{2}}{\partial z^{2}}\!+\!V_{\mathrm{R}}(z)\!+\!iV_{\mathrm{I}%
}(z)\!+\!g_{\mathrm{1D}}|\Psi |^{2}\right] \Psi ,  \label{gpe}
\end{equation}%
here the complex amplitude distribution $\Psi \equiv \Psi (z,t)$ is
macroscopic order parameter, the nonlinearity $g_{\mathrm{1D}}=2a_{s}\hbar
\omega _{\bot }$ is the effective 1D coupling strength, where $a_{s}$ stands
for the $s$-wave scattering length which can be changed through Feshbach
resonance, and $\omega _{\bot }$ denotes the transverse confining frequency.
The external potential is chosen as a harmonic trap $V_{\mathrm{R}%
}(z)=(1/2)m\omega _{z}^{2}z^{2}$ with $m$ being the atomic mass and $\omega
_{z}$ is the longitudinal confining frequency~\cite{BEC}. The gain/loss
potential $V_{\mathrm{I}}(z)$ is phenomenologically incorporated to account
for the manipulation of gain/loss potential in BECs by a novel experimental
microscopy technique \cite{Dis}.

After the dimensionless transformations of the density, length, time and
energy measured in units of $g/(2a_{s}),\,a_{\bot }=\sqrt{{\hbar }/(m\omega
_{\bot })},\,\omega _{\bot }^{-1}$, and $\hbar \omega _{\bot }$, we arrive
at an effective 1D GP equation with non-Hermitian $\mathcal{PT}$-symmetric
potential in the dimensionless variables
\begin{equation}
i\frac{\partial \psi }{\partial t}=\!\left[ -\frac{1}{2}\frac{\partial ^{2}}{%
\partial z^{2}}+v_{\mathrm{R}}(z)+iv_{\mathrm{I}}(z)\,+g|\psi |^{2}\right]
\psi  \label{nls}
\end{equation}%
which is associated with $\delta \mathcal{L}/\delta \psi ^{\ast }=0$ in
which the Lagrangian density can be written as $\mathcal{L}=i(\psi \psi
_{t}^{\ast }-\psi ^{\ast }\psi _{t})+|\psi _{z}|^{2}+2[v_{\mathrm{R}}(z)+iv_{%
\mathrm{I}}(z)]|\psi |^{2}+g|\psi |^{4}$, where $v_{\mathrm{R}%
}(z)=(1/2)\omega ^{2}z^{2}$ with the trap parameter $\omega =\omega
_{z}/\omega _{\bot }$, $v_{\mathrm{I}}(z)=\hbar \omega _{\bot }V_{\mathrm{I}%
}(z)$, and $g$ is a interaction parameter. Based on the previous discussion,
the real and the imaginary components of the $\mathcal{PT}$-symmetric
potential should satisfy the relations: $v_{\mathrm{R}}(z)=v_{\mathrm{R}%
}(-z) $ and $v_{\mathrm{I}}(-z)=-v_{\mathrm{I}}(-z)$, respectively. We here
concentrate on the attractive interaction $g<0$. The repulsive interaction $%
g>0$ does not pose new changes and will be considered elsewhere.

We here focus on spatially localized soliton solutions of Eq.~(\ref{nls})
for which $\lim_{|z|\rightarrow \infty }\psi (z,t)=0$. Our goal is to seek a
class of soliton-type stationary solutions in the form $\psi (z,t)=\phi
(z)\exp (-i\mu t)$ with $\mu $ being the chemical potential and the \textit{%
complex function} $\phi (z)\in \mathbb{C}[z]$ obeying the stationary GP
equation with $\mathcal{PT}$-symmetric potential
\begin{equation}
\mu \phi =\left[ -\frac{1}{2}\frac{d^{2}}{dz^{2}}+v_{\mathrm{R}}(z)+iv_{%
\mathrm{I}}(z)\,+g|\phi |^{2}\right] \phi .  \label{pode}
\end{equation}

Let us now consider the more interesting and physically relevant $\mathcal{PT%
}$-symmetric potential with the external trapping potential $v_{\mathrm{R}%
}(z)$ and the gain/loss potential $v_{\mathrm{I}}(z)$ being of the harmonic
trap $v_{\mathrm{R}}(z)=(1/2)\omega ^{2}z^{2}$ and a family of the
Hermite-Gaussian distributions
\begin{equation}
v_{\mathrm{I},n}(z)\!\!=\!w_{0}\!\sqrt{\omega }\,\left[ 2nH_{n\!-\!1}(\!%
\sqrt{\omega }z)\!\!-\!\!\sqrt{\omega }zH_{n}(\!\sqrt{\omega }z)\right]
\!e^{\!-\!\frac{\omega z^{2}}{2}},  \label{gamma}
\end{equation}%
where $w_{0}$ is a constant parameter which can be controlled
experimentally, the even integer index $n$ (i.e., $n=0,2,4,...$) is
necessary to make sure $v_{\mathrm{I},n}(z)$ be the odd function for the $%
\mathcal{PT}$-symmetric potential and determines the number of nodes across
the Gaussian envelope, $H_{n}(z)=(-1)^{n}e^{z^{2}}d^{n}e^{-z^{2}}/dz^{n}$
stands for the Hermite polynomial. For the odd integer index $n$, the
function $v_{\mathrm{I},n}(z)$ given by Eq.~(\ref{gamma}) does not obey the $%
\mathcal{PT}$-symmetric condition, so we do not considered these cases here.

As a consequence, we find that there exists a family of exact $\mathcal{PT}$%
-symmetric soliton-type solutions as
\begin{equation}
\phi _{n}(z)\!=\frac{\sqrt{2}w_{0}}{3\sqrt{|g|}}H_{n}(\!\sqrt{\omega }%
z)e^{-\!\frac{\omega z^{2}}{2}}\exp [i\varphi _{n}(z)],  \label{1dsolution}
\end{equation}%
where $n=0,2,4,...,\,g<0$, the chemical potential satisfies the condition $%
\mu =\omega \lbrack n+1/2]$, and the phase $\varphi _{n}(z)$ is given by $%
\varphi _{n}(z)=(2/3)w_{0}\int_{0}^{z}[H_{n}(\!\sqrt{\omega }s)\exp (-\omega
s^{2}/2)]ds.$ It is easy to show that these solutions is well localized
since $\left. \phi _{n}(z)\right\vert _{|z|\rightarrow \infty }\rightarrow
0. $

In the following, without loss of generality, we will study the cases of $%
n=0 $ and $n=2$, respectively, for their spontaneous $\mathcal{PT}$ symmetry
breaking, stabilities and collisions properties. Thus, we can write the
gain/loss terms in the forms of
\begin{subequations}
\begin{eqnarray}
&&v_{\mathrm{I},0}(z)=-w_{0}\omega z\exp \left( -\omega z^{2}/2\right) ,
\label{n0} \\
&&v_{\mathrm{I},2}(z)=w_{0}\omega (10z-4\omega z^{3})\exp \left( -\omega
z^{2}/2\right) ,\quad  \label{n2}
\end{eqnarray}%
for $n=0$ and $n=2$, respectively. The corresponding exact $\mathcal{PT}$%
-symmetric solitons should be written as
\end{subequations}
\begin{subequations}
\begin{eqnarray}
&&\phi _{0}(z)\!=\!\!\frac{\sqrt{2}w_{0}}{3\sqrt{|g|}}e^{-\!\frac{\omega
z^{2}}{2}}\!\exp \!\left[ \frac{iw_{0}}{3}\sqrt{\frac{2\pi }{\omega }}\,%
\mathrm{erf}\left( \sqrt{\omega /2}z\right) \!\right] ,\quad  \label{n0_S} \\
&&\phi _{2}(z)\!=\frac{2\sqrt{2}w_{0}}{3\sqrt{|g|}}(2\omega z^{2}-1)e^{-\!%
\frac{\omega z^{2}}{2}}\exp [i\varphi _{2}(z)],  \label{n2_S}
\end{eqnarray}%
for $n=0$ and $n=2$, respectively, where $\mathrm{erf}\left( x\right) $
denotes error function and $\varphi _{2}(z)=(2/3)w_{0}[\sqrt{2\pi /\omega }\,%
\mathrm{erf}(\sqrt{\omega /2}\,z)-4z\exp (-\omega z^{2}/2)]$.

\begin{figure}[tbp]
\begin{center}
\vspace{-0.1in} {\scalebox{0.35}[0.35]{\includegraphics{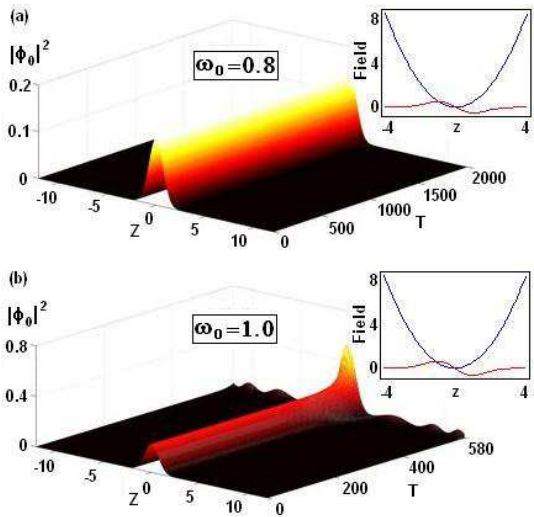}}}
\end{center}
\par
\vspace{-0.1in}
\caption{(color online). Intensity evolution of a single-peak soliton state
in a \textit{$\mathcal{PT}$-}symmetric\textit{\ }potential (\protect\ref{n0}%
) for $g=-1$ and $\protect\omega =1$. (a) $w_{0}=0.8$. (b) $w_{0}=1.0$. The
inset depicts the real (top) and imaginary (bottom) component of $\mathcal{PT%
}$- symmetric potential.}
\label{fig1}
\end{figure}

As is well known that, for our studied \textit{$\mathcal{PT}$-}symmetric%
\textit{\ }potentials, the linear spectrum can be entirely real as long as
the $w_{0}$ is operated below the phase transition point (unbroken \textit{$%
\mathcal{PT}$} symmetry) for a fixed value of $\omega $. Above this
so-called \textit{$\mathcal{PT}$-}symmetric threshold, a phase transition
occurs and the linear spectrum enters the complex domain.

Notice that in our studied system, the corresponding linear spectrum problem
associated with the \textit{$\mathcal{PT}$-}symmetric potential (\ref{n0}),
exhibits an entirely real spectrum provided that,
\end{subequations}
\begin{equation}
\left\vert w_{0}\right\vert \leq 1.945,  \label{w0_n0}
\end{equation}%
for the case of $g=-1$, $\omega =1$ and $n=0$. A single-peak soliton-type
stationary state (\ref{n0_S}) in the \textit{$\mathcal{PT}$-}symmetric
potential (\ref{n0}) for the case of $g=-1$, $\omega =1$ and $w_{0}=0.8$ is
shown in Fig.~\ref{fig1}(a). The stability of this $\mathcal{PT}$-symmetric
soliton has been confirmed by numerical simulation of Eq.~(\ref{nls}) using
beam propagation methods with adding random noise on both amplitude and
phase of exact $\mathcal{PT}$-symmetric soliton state. Thus for this given $%
w_{0}$, one can see that \textit{$\mathcal{PT}$-}symmetric\textit{\ }soliton
states can be found with real energy in this system. Very interestingly, one
finds that even if the potential (\ref{n0}) has below the phase transition
point ($w_{0}=1.0,$ its single particle spectrum is real), single-peak
soliton has already been unstable with breaking \textit{$\mathcal{PT}$}
symmetry and\textit{\ }imaginary energy. Fig.~\ref{fig1}(b) shows the
numerical simulation of Eq.~(\ref{nls}) for the case of $g=-1$, $\omega =1$
and $w_{0}=1.0$ by using the same method as in Fig.~\ref{fig1}(a). This
means that, the single-peak soliton itself can modify the real part of
potential through the mean field interaction term in Eq.~(\ref{nls}). Thus,
this new effective trapping potential shifts the \textit{$\mathcal{PT}$ }%
threshold $w_{0}$ of Eq. (\ref{w0_n0}) towards a smaller value and in turn
damages this stationary soliton state\ against breaking \textit{$\mathcal{PT}
$} symmetry. For the obtained single-peak $\mathcal{PT}$-symmetric soliton (%
\ref{n0_S}) we find that the quantity $S=(i/2)(\phi\phi_x^{*}-\phi_x%
\phi^{*})=4\omega_0^3/(9|g|)\exp(-3\omega z^2/2)$ associated with the
transverse power flow density is positive with $\omega_0>0$, which means
that the power always flows in one direction, i.e., from the gain to loss
region.

\begin{figure}[tbp]
\begin{center}
\vspace{-0.1in} {\scalebox{0.35}[0.35]{\includegraphics{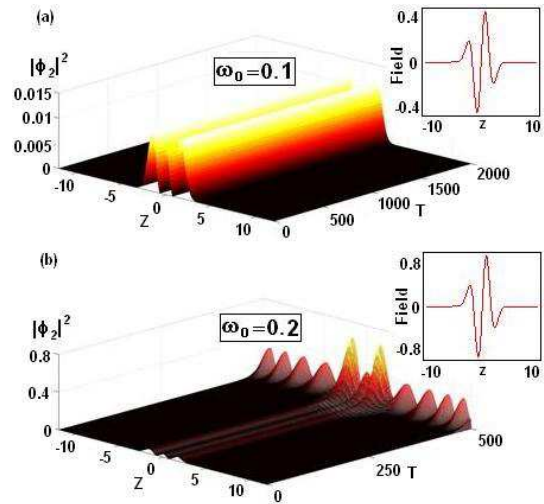}}}
\end{center}
\par
\vspace{-0.1in}
\caption{(color online). Intensity evolution of a multi-peak soliton state
in a \textit{$\mathcal{PT}$-}symmetric\textit{\ }potential (\protect\ref{n2}%
) for $g=-1$ and $\protect\omega =1$. (a) $w_{0}=0.1$. (b) $w_{0}=0.2$. The
inset depicts the imaginary component of $\mathcal{PT}$- symmetric
potential. }
\label{fig2}
\end{figure}

We next investigate multi-peak soliton-type stationary state (\ref{n2_S})
and their dynamics supported with the \textit{$\mathcal{PT}$-}symmetric
potential (\ref{n2}). The linear spectrum properties of such a \textit{$%
\mathcal{PT}$-}symmetric potential can be understood by examining the
corresponding linear spectrum problem of Eq.~(\ref{nls}), And one finds that
purely real spectrums are possible in the range
\begin{equation}
\left\vert w_{0}\right\vert \leq 0.00002,  \label{w0_n2}
\end{equation}%
for the case of $g=-1$\text{, }$\omega =1$ and $n=2.$ Since the similar idea
holds for such \textit{$\mathcal{PT}$-}symmetric potential (\ref{n2}), by
numerical simulation of Eq.~(\ref{nls}) using beam propagation methods,
however, one finds a stable \textit{$\mathcal{PT}$-}symmetric multi-peak
soliton-type state even for the parameters of $g=-1$\text{, }$\omega =1$ and
$w_{0}=0.1$ as is shown in Fig.~\ref{fig2}(a). Thus, very interestingly, in
contrast to the stable \textit{$\mathcal{PT}$-}symmetric single-peak
soliton, here, the multi-peak soliton itself eventually, shifts the \textit{$%
\mathcal{PT}$ }threshold $w_{0}$ of Eq.~(\ref{w0_n2}) towards a rather
larger value and makes soliton-type state\ preserve \textit{$\mathcal{PT}$}
symmetry very well by drastically modifying the real part of potential
through the mean field interaction term. With increasing the $w_{0}$
further, the multi-peak soliton will be unstable with breaking \textit{$%
\mathcal{PT}$} symmetry and\textit{\ }imaginary energy. Fig.~\ref{fig2}(b)
shows the numerical simulation of Eq.~(\ref{nls}) for the case of $g=-1$%
\text{, }$\omega =1$ and $w_{0}=0.2$ by using the same method as former
cases.

\begin{figure}[!ht]
\begin{center}
\vspace{0in} {\scalebox{0.42}[0.5]{\includegraphics{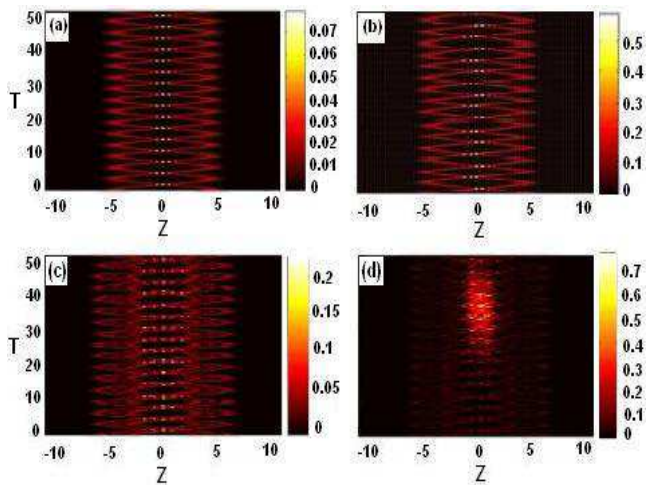}}}
\end{center}
\par
\vspace{-0.15in}
\caption{(color online). Dynamic evolution of collisions of two single-peak
and multi-peak solitons in a \textit{$\mathcal{PT}$-}symmetric\textit{\ }%
potentials (\protect\ref{n0}) and (\protect\ref{n2}) for $g=-1$ and $\protect%
\omega =1$. Here (a) $w_{0}=0.3$ and (b) $w_{0}=0.8$ for two single-peak
solitons. (c) $w_{0}=0.2$ and (d) $w_{0}=0.25$ for two multi-peak solitons. }
\label{fig3}
\end{figure}

Finally, we need to check if this \textit{$\mathcal{PT}$}-symmetric soliton
indeed behave as solitons. To this end, we take initial conditions in the
form of a superposition of two single-peak (\ref{n0_S}) or multi-peak (\ref%
{n2_S}) solitons with symmetric central positions. In such a case, and given
that the single-peak or multi-peak solitons were found above to be robust
objects behaving similarly to solitons of an integrable system, one may
expect that the solitons would perform harmonic oscillations in the presence
of the sufficiently large parabolic trap even with $\mathcal{PT}$-symmetric
potentials. As shown in the Fig.~\ref{fig3}(a) for the case of $g=-1$, $%
\omega =1$ and $w_{0}=0.3$, two single-peak solitons propagate in opposite
directions and undergo a pretty elastic collision and the two solitons
remain unscathed after the collision. The stability of this superposition of
two single-peak solitons has been confirmed by numerical simulation of Eq.~(%
\ref{nls}) using beam propagation methods as before. But when the gain/loss
parameter $w_{0}\ $becomes larger (for example, $w_{0}=0.8$), two
single-peak solitons undergo a inelastic collision as shown in the Fig.~\ref%
{fig3}(b). A superposition of two multi-peak solitons (\ref{n2_S}) with
symmetric central positions is also found to be robust objects behaving
similarly to above and undergos a elastic collision as shown in the Fig.~\ref%
{fig3}(c) for the case of $g=-1$, $\omega =1$ and $w_{0}=0.2$. But when the
gain/loss parameter $w_{0}\ $becomes a little larger (for example, $%
w_{0}=0.25$), two multi-peak solitons become unstable clearly as shown in
the Fig.~\ref{fig3}(d). All these findings certainly indicate that many
kinds of robust stable nonlinear modes can be supported by \textit{$\mathcal{%
PT}$}-symmetric potential in BECs.

Experimental realization of the ideas presented here is promising. A
condensate in a dilute vapor of sodium or rubidium atoms is placed in a
cigar-shaped potential may provide an accessible laboratory to
experimentally observe spontaneous \textit{$\mathcal{PT}$} symmetry breaking
and stability of solitons. Such an experiment would require precise control
over the rates of gain/loss process in the condensate where atoms are
injected continuously into the condensate in left side, whereas they are
removed simultaneously from the condensate in right side. The combination of
high spatial resolution with \textit{in situ }detection \cite{Dis} will
hopefully provide new possibilities for the preparation, the manipulation
and the characterization of non-Hermitian $\mathcal{PT}$ symmetry condensate.

In summary, we have demonstrated that a novel class of exact $\mathcal{PT}$%
-symmetric solitons can be well supported in BECs with $\mathcal{PT}$%
-symmetric gain/loss potential. Their spontaneous $\mathcal{PT}$ symmetry
breaking, stability, as well as collision dynamics of such \textit{$\mathcal{%
PT}$}-symmetric solitons were examined in detail. We observe the significant
effects of mean-field potential by modifying the threshold point of
spontaneous \textit{$\mathcal{PT}$} symmetry breaking. Our results may
intrigue a new class of \textit{$\mathcal{PT}$}-synthetic universal
properties study that relies on manipulations of macroscopic matter-wave
field with BECs.

\acknowledgments

This work was supported by the NSFC under grants Nos. 60821001/F02,
10874235, 10934010, 60978019, the NKBRSFC under grants Nos. 2009CB930701,
2010CB922904 and 2011CB921500.

\end{document}